# Two Dimensional Random Patterns

Chakradhara Reddy Chinthapanti
Computer Science Department
Oklahoma State University, Stillwater, OK – 74078

**ABSTRACT**

A new approach to the generation of random sequences and two dimensional random patterns is proposed in this paper in which random sequences are generated by making use of either Delaunay triangulation or Voronoi diagrams drawn from random points taken in a two dimensional plane. Both the random sequences and two dimensional random patterns generated in this manner are shown to be more random when compared to pseudo-random sequences and patterns.

**INTRODUCTION**

A pseudo-random binary sequence (PN sequence), a sequence of binary digits of length $2^k-1$, is also called a maximal-length shift-register sequence [18]. PN sequences are widely used in cryptography, signal design, scrambling, fault detections and simulations (e.g., Monte Carlo methods) [9]. A decimal sequence (d-sequence) is nothing but a decimal expansion of a rational number which is represented in a sequence of digits. The d-sequence generated from a particular number can be periodic, irregular or it may terminate. The papers by Kak [1] - [6] give the theory of d-sequences and their properties.

A pseudo random sequence can be represented as a d-sequence of a rational number. Hence, d-sequences are also used in the place of pseudo random sequences. A binary d-sequence is generated by using the formula, $a(i) = 2^i \bmod p \bmod 2$, where p is a prime number and a maximum-length sequence is generated with period p-1, when 2 is a primitive root of p. These sequences are periodic and therefore examining the sequence in one period is enough to get to know the random properties. D-sequences cannot be used directly in computationally secure random number generator (RNG) applications. The reason being it is easy to find i given $\log_2 p$ bits of a(i) [2]. Hence by adding two or more different d-sequences (obtained by using primes p1, p2…) mod 2, non-linearity is introduced in the generation of random sequence [6]. The resulting sequence is a better option to be used as random sequence than the previous one.

Coming back to the problem of randomness, pseudo random sequences and decimal sequences appear to be random but they are completely deterministic. The physical nature of randomness is discussed in [19]-[22]. The question of randomness defined in terms of the complexity of the algorithm needed to generate it is given in [23]. Random sequences for scrambling and for two-dimensional patterns are given in [7],[8].

The pseudo-random sequence appears to be random in the sense that the binary values and groups or runs of the same binary value occur in the sequence in the same proportion they would if the sequence were being generated based on a fair "coin tossing" experiment. In the experiment, each head could result in one binary value and a tail the other value. The PN sequence appears to have been generated from such an experiment. The sequence is not truly random in that it is completely determined by a relatively small set of initial values. It was also shown in [2] that it is easy to find i given $\log_2 p$ bits of a(i), therefore, d-sequences cannot be directly used in RNG applications. The further studies tried to increase the period but still the decimal sequence would be generated from the prime reciprocals. This paper presents a new approach to generate the random sequence from the Delaunay triangulation or Voronoi diagram drawn from the random points which would be more random in nature. The randomness is shown by performing autocorrelation function and diehard tests.

When the studies were going on the one dimensional random sequences, some applications called for the two dimensional (2D) sequences. Two dimensional arrays of area $2^{lk}-1$ are discovered long back. Two dimensional random arrays and patterns have applications in a variety of areas including scrambling and fault detection. MacWilliams and Sloane [9] considered a n1 × n2 array, where n1 and n2 are relatively prime numbers and they used pseudo-random sequences to produce the two dimensional array. Two dimensional patterns are basic to visual perception and it is not known how exactly such patterns are coded and recalled [15], although there is evidence that coding is unary in certain situations for one-dimensional patterns [16],[17]. We can also replace the pseudo-random sequences by the prime reciprocal sequences which gives the same results. By using the idea of prime reciprocals, generalization to two dimensional patterns by using two dimensional polynomials rather than primes is also studied [8].

In our approach, the random sequences generated from the Delaunay triangulation or Voronoi diagram drawn from the random points are used to generate the 2D random patterns. It is proposed that even the 2D pattern would be more randomized when it is generated from this approach. The 2D patterns can be generated in various styles using any number of random sequences. This is discussed later in the paper. The autocorrelation tests and diehard tests [14] also proves the randomness of the pattern.

**DELAUNAY TRIANGULATION AND VORONOI DIAGRAMS**

We begin with a consideration of Delaunay triangulation and Voronoi diagrams [10] - [13]. In brief, a Voronoi diagram is a special kind of decomposition of a metric space determined by distances to a specified discrete set of objects in the space, e.g., by a discrete set of points. It is obtained by drawing the bisectors for the two nearest points and the vertex of the generated polygon would be the center of the circle on which the corresponding closest points would be lying. In our approach, the edges which have their length to infinity and which have their intersection outside the working plane would be restricted to the working plane by clipping the polygons accordingly as shown in figure 1.

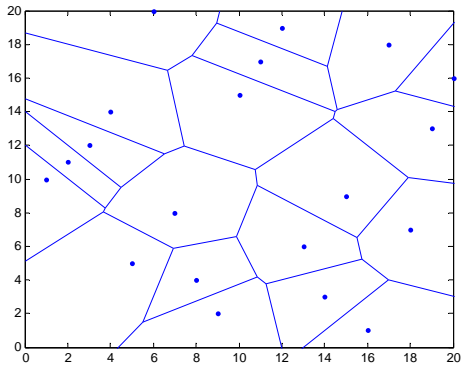

Figure 1. Clipped voronoi diagram for 20 random points (working plane is (0, 0, 20, 20)).

A Delaunay triangulation for a set of points in the plane is a triangulation such that no point in the set of points considered would lie inside the circumcircle of any triangle in the triangulation. Delaunay triangulations maximize the minimum angle of all the angles of the triangles in the triangulation and they tend to avoid skinny triangles. The Voronoi diagram and Delaunay triangulation are called duals as one can be generated from the other. A Delaunay triangulation generated for 20 random points is shown in figure 2.

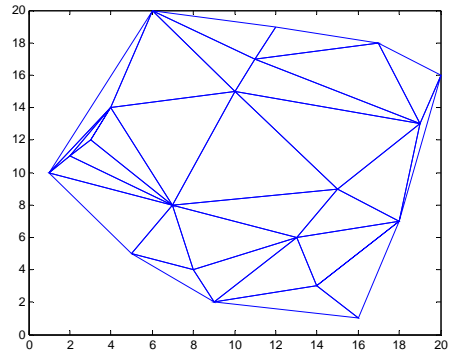

Figure 2. Delaunay triangulation for 20 random points.

**PROPOSED APPROACH TO GENERATE RANDOM SEQUENCES**

- Generate both the X and Y co-ordinates of the points using a RNG.
- For the generated random points, Delaunay triangulation or Voronoi diagram is generated.
- In Delaunay triangulation, the mean area of all the triangles is calculated and for each individual triangle, if its area is more than the mean area, 1 is considered and if it is less than the mean area, 0 is considered.
- If a Voronoi diagram is chosen, the edges which have their length to infinity are restricted to the working plane and clipped polygons are considered. Now, the mean area of all the polygons is calculated and if the individual polygon area is more than the mean area, 1 is considered and if it is less than the mean area, 0 is considered.
- The sequence is generated with these 1's and 0's in the same order as the individual triangles or polygons (their areas), as they are considered for the comparison with the mean area.

The binary sequence generated from the above procedure is a better candidate than the previous random sequences. The unpredictability is more, as one cannot predict the next digit in the generated sequence at any length which is highly unlikely compared to d-sequences or pseudo-random sequences. Before going to results, there is one more point to be noted. The main idea is to present a new approach to generate random sequences and not on the security issues. Also, there will be no Delaunay triangulation when the random points are on a straight line and no Voronoi diagram when the random points are either on a straight line or on a circle. But it is almost impossible for such a condition to occur as the X and Y co-ordinates would be generated by a RNG. The number of polygons (along with clipped) generated from the Voronoi diagram would be same as the number of random points taken (shown in figure 3, on left side). Whereas, the number of triangles generated from the Delaunay triangulation would be 2n-k-2, where n is the number of random points and k is the number of points on the convex hull of the given set of points (shown in figure 3, on right side).

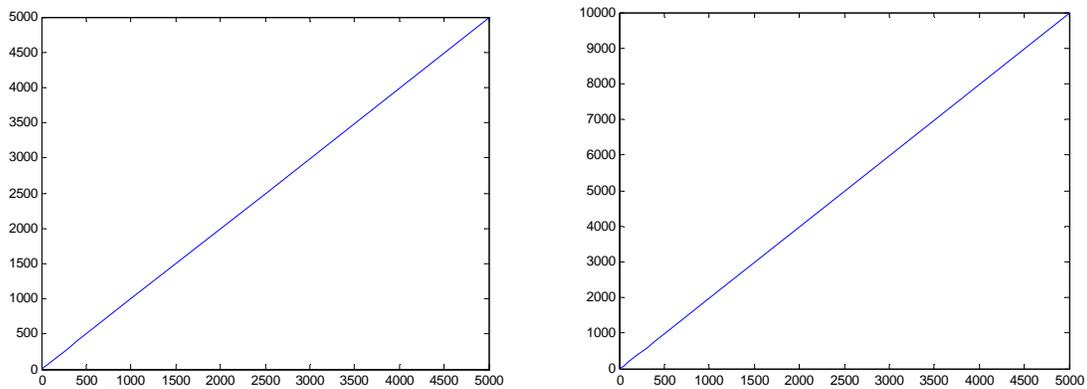

Figure3. Number of random points (X-axis) Vs Number of polygons (left) / triangles (right) generated from the Voronoi diagram/ Delaunay triangulation (Y-axis).

**EXAMPLES**

The simulations are done in MATLAB V 7.10 (R 2010a) and the results are shown only for the random sequences generated from the Delaunay triangulation. Figure 4 shows the Delaunay triangulation generated for 10, 100, 1000 and 5000 random points and the binary sequences of length 12, 184, 1982 and 9977 are generated respectively. The sequence generated for 10 random points is 010111110011.

The sequence generated for 100 random points is 0001100010000001110110111110001100100100011000000001100000001111100100000001101001011100000101011000010100100000100101111000000010001000011101100111100110000011101101000000000. The random sequences generated for 1000 and 5000 random points are

not shown here (as it would consume the whole paper). Unlike d-sequences, one cannot predict the next coming bits or digits when some or all the previous bits of the sequence are known.

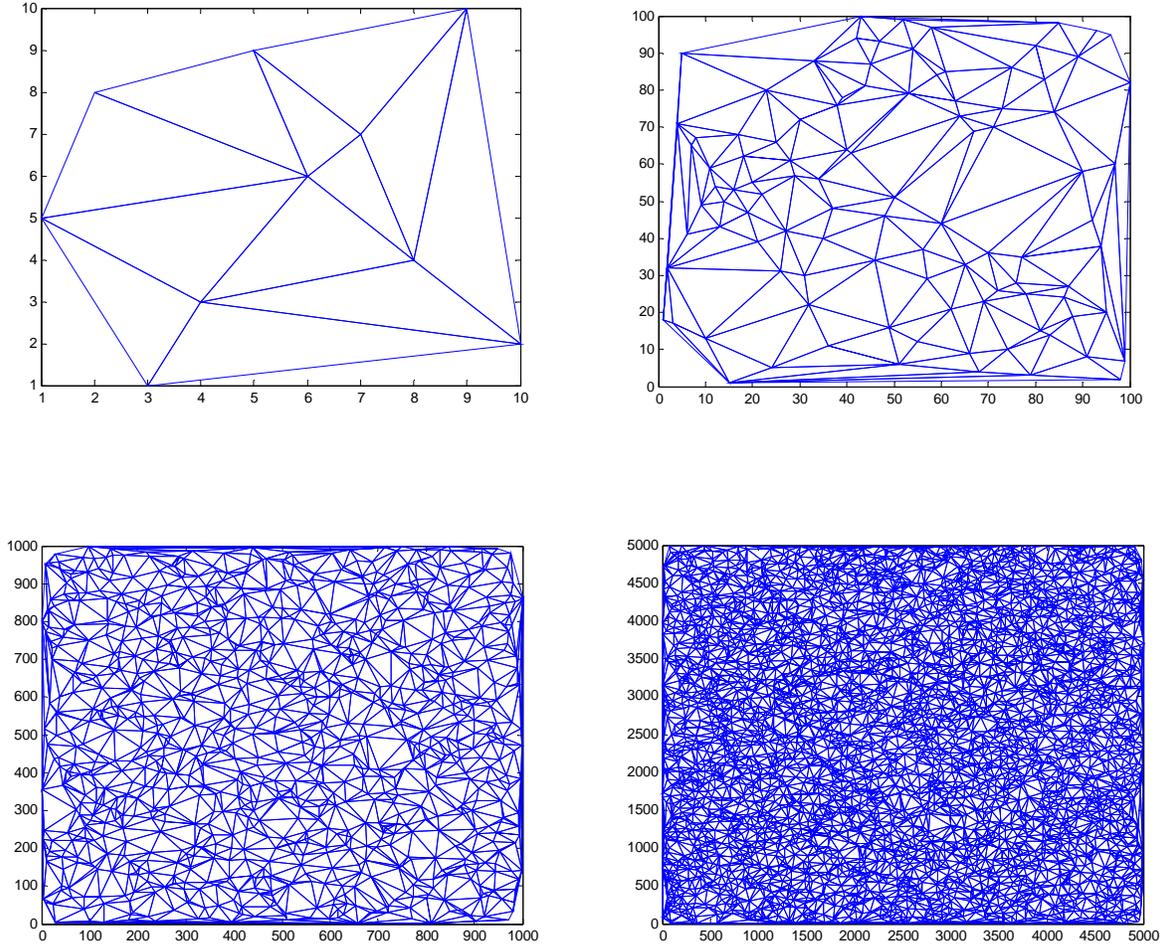

Figure 4. Delaunay triangulation for 10 (top left), 100 (top right), 1000 (bottom left) and 5000 (bottom right) random points

## AUTOCORRELATION FUNCTION

The autocorrelation function for a binary maximum length decimal sequence is given as,

$$C(j) = (1/k)\sum_{i=1}^{k} a_i a_{i+j} ,$$

where k is the maximum length, j= 0, 1, 2, 3…. In the binary sequence of 1's and 0's, 0 is replaced by -1 and the test is carried out. The graph should be a symmetric one with a maximum value of 1 for j=0, k and all the remaining values being 0.

The graph in figure 5 (left) is the autocorrelation function of the random sequence of length 184 (100 random points). The maximum value of 1 is obtained for 0 and 184, as it should be. The intermediate values are between +0.2 and -0.2.

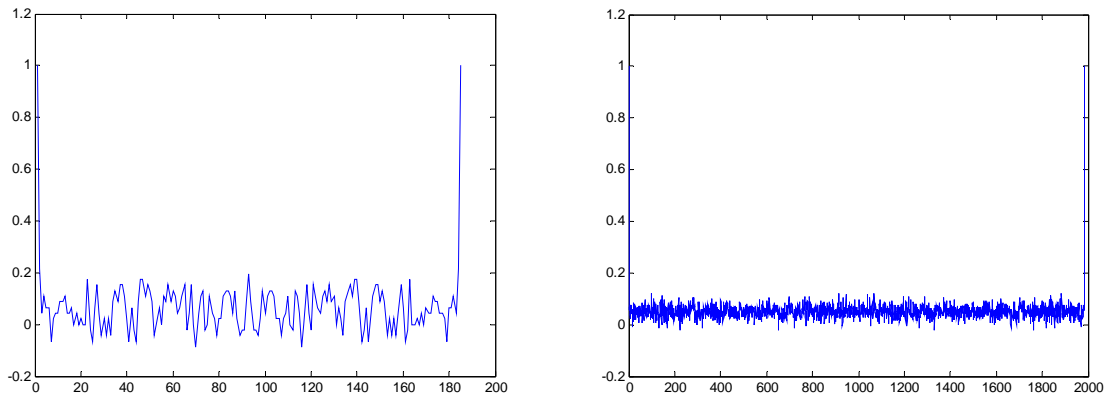

Figure 5. Autocorrelation function for random sequence of length 184 (left) and for random sequence of length 1982 (right)

As we consider the autocorrelation function of the random sequence of length 1982 (1000 random points), the intermediate values lie between +0.1 and + 0.1, as shown in figure 5 (right). Thus, when the length of the sequence increases, the ideal two-valued autocorrelation function would be met.

**DIEHARD TESTS RESULTS**

The diehard tests proposed by George Marsaglia are a series of statistical tests for measuring the quality of a random number generator [14]. These tests can be used to prove the randomness of these random sequences. The diehard tests are simulated in JAVA by Zur Aougav under sourceforge (open source software), and his test suite, jrandtest-0.4, is used to test the random sequence. The test suite also provides diehard tests for so many algorithms.

Table 1 provides the diehard tests results of the random sequence and a comparison is provided with other random number generators namely SHA1 algorithm, JAVA random function. In our simulation, all the tests which follow a distribution are using chi-square (KS) test.

| Test | JAVA Random | SHA1 Random | Random sequence |
|---|---|---|---|
| Runs test (sequence of 10000) | RunsUP: KS test for 10 p's: 0.127  Runs DOWN: KS test for 10 p's: 0.698 | RunsUP: KS test for 10 p's: 0.132  Runs DOWN: KS test for 10 p's: 0.751 | RunsUP: KS test for 10 p's: 0.136  Runs DOWN: KS test for 10 p's: 0.798 |
| Squeeze test | Chi-square with 42 degrees of freedom: 52.6387  z-score =1.1608,  p-value = 0.1258 | Chi-square with 42 degrees of freedom: 42.34  z-score = 0.0371,  p-value = 0.4563 | Chi-square with 42 degrees of freedom: 38.19  z-score = 0.145,  p-value = 0.6 |
| Min distance (done for 100 sets) | KS test on 100 transformed mindist^2's:  p-value = 0.2174 | KS test on 100 transformed mindist^2's:  p-value = 0.2110 | KS test on 100 transformed mindist^2's:  p-value = 0.3735 |
| Count the 1's (sample size: 256000) | Chisquare = 944,003.9000  z-score = 13,314.8758  p-value = 0.0000 | Chisquare = 942,790.6199  z-score = 13,297.7175  p-value = 0.0000 | Chisquare = 963,889.3220  z-score = 13,596.0982  p-value = 0.0000 |
| Birthday spacings (bdays=1024, days/yr=2^24, lambda=16, sample size=500) | degree of freedoms : 17  p-value for KStest on the 9 p-values: 0.0894 | degree of freedoms : 17  p-value for KStest on the 9 p-values: 0.3454 | degree of freedoms : 17  p-value for KStest on the 9 p-values: 0.2378 |
| Binary rank (32x32 matrices) | chi-square =4.3138 with df = 3;  p-value = 0.2077 | chi-square =8.0225 with df = 3;  p-value = 0.0409 | chi-square =7.0140 with df = 3;  p-value = 0.0642 |

Table 1. Die hard test results for the random sequence, SHA1 algorithm and JAVA random function.

There are also other diehard tests like parking lot test, monkey test and overlapping sums test whose results are not provided in the table. These tests are also following the KS test. Hence, all the above results suggest that the random sequence generated from the Delaunay triangulation or Voronoi diagram can be considered as equally random as the other RNGs.

**TWO DIMENSIONAL RANDOM PATTERNS USING RANDOM SEQUENCES**

The two dimensional random patterns can be generated from either d-sequences or random sequences generated from the above proposed method. As we know, the decimal sequence of a prime reciprocal, 1/p, is generated using the formula $2^i$ mod p mod 2. When the sequence is placed into rows and columns (any order can be used), a 2D random pattern of desired size can be generated. By converting the 1's and 0's to black and white pixels, even an image of desired size can be generated.

The two dimensional random patterns are generated from the random sequences which are generated from the Delaunay triangulation or Voronoi diagram in the same way as mentioned above. As the quality of these random sequences is already proved in the above tests, the 2D random patterns generated from these sequences can also be considered as more randomized than the other patterns.

**EXAMPLES**

A 128 × 64 image generated from a single random sequence obtained from the Delaunay triangulation is shown in figure 6 (top left). Here the length of the sequence should be 128 × 64 = 8192. Hence, a random sequence with a maximal length of 8192 (generated from 4108 random points) is generated from the above approach. Figure 6 (top center) shows the 128 × 64 image generated from four different random sequences of length 1982 (1000 random points), 3971 (2000 random points), 982 (500 random points) and 1257 (637 random points).

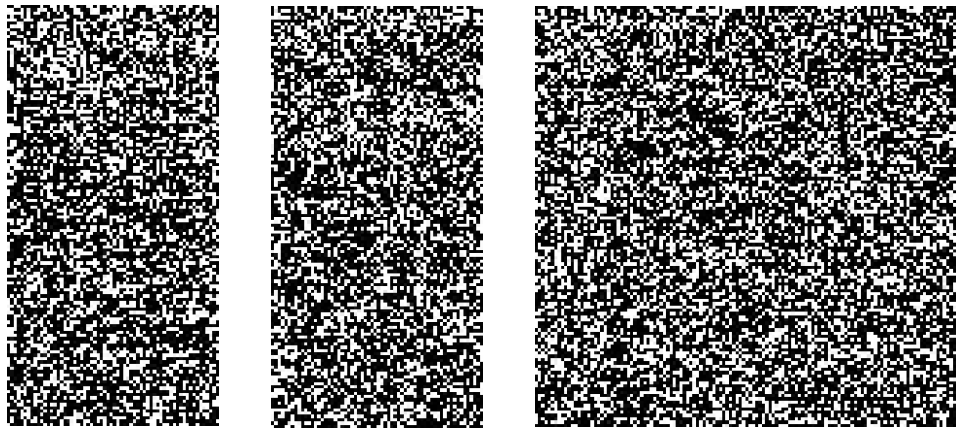

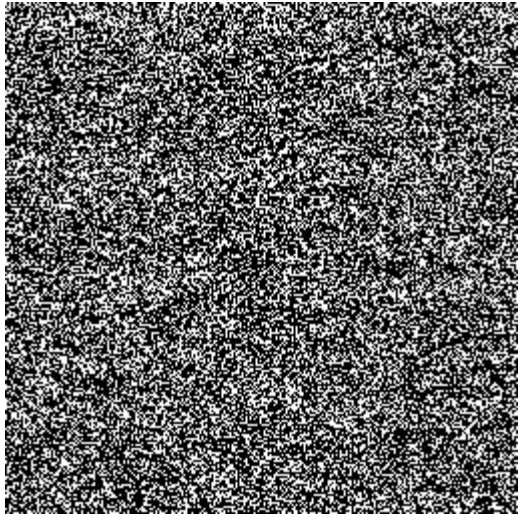

Figure 6. Images of different sizes generated using one or more random sequences.

Figure 6 (top right) shows a 128 × 128 image (16284 pixels) generated from seven different random sequences of length 4976 (2500 random points), 583 (300 random points), 1379 (700 random points), 2376 (1200 random points), 984 (500 random points), 2979 (1500 random points) and 3107 (1564 random points). Figure 6 (bottom) shows a 256 × 256 image (65536 pixels) generated from six different random sequences of length 19973 (10000 random points), 14977 (7500 random points), 9978 (5000 random points), 5975 (3000 random points), 11972 (6000 random points) and 2661 (1343 random points).

**TWO DIMENSIONAL AUTOCORRELATION FUNCTION**

The two dimensional autocorrelation function for a binary maximum length decimal sequence is given as,

$$C(i,j) = (1/k1)(1/k2)\sum_{m=1}^{k1}\sum_{n=1}^{k2} a(m,n).a(m+i, n+j)　,$$

where k1 is the row size and k2 is the column size, i and j take values 0, 1, 2, 3…. In the binary sequence of 1's and 0's, 0 is replaced by -1 and the test is carried out. The graph should be a symmetric one with a maximum value of 1 for j=0, k (here k=k1*k2) and all the remaining values being 0.

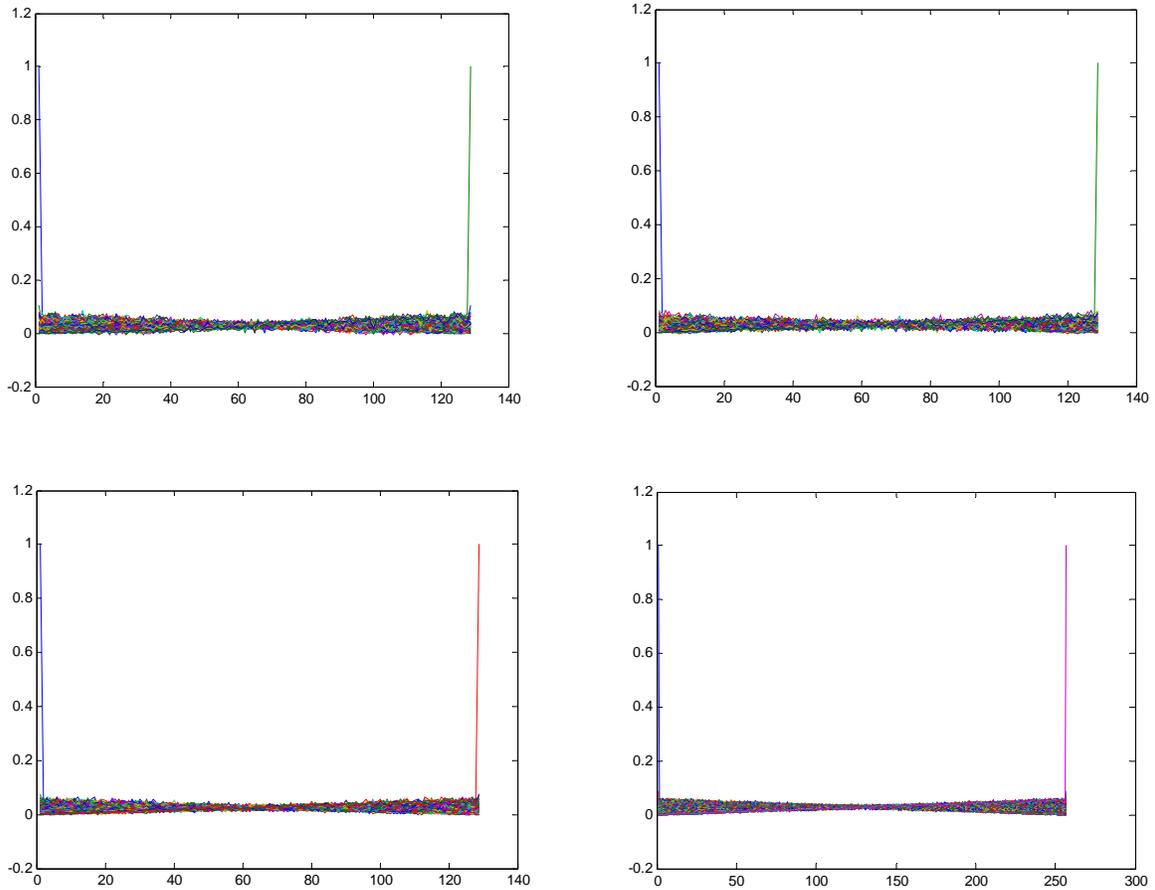

Figure 7. The 2D autocorrelation function for a 128×64 image (top left), 128×64 image (top right), 128×128 image (bottom left) and 256×256 image (bottom right).

Figure 7 shows the 2D autocorrelation function results for the images generated in figure 6 using different random sequences. In all the graphs, maximum value of 1 is obtained for 0 and k. But, the intermediate values lie between +0.1 and -0.1.

**CONCLUSION**

This paper presents a new approach to the generation of random sequences and generation of the two dimensional random patterns. These random sequences are more random in nature compared to the d-sequences and PN sequences as given by the autocorrelation and diehard tests. Hence, effectively randomized 2D patterns could be generated from these sequences in various ways.